# Set-based Complexity and Biological Information

David J. Galas*[1,2], Matti Nykter[1,3], Gregory W. Carter[1],
Nathan D. Price[1+] and Ilya Shmulevich[1]

[1]Institute for Systems Biology, Seattle, WA, USA
[2]Battelle Memorial Institute, Columbus, OH, USA
[3] Institute of Signal Processing, Tampere University of
Technology, Tampere, Finland

## Summary

It is not obvious what fraction of all the potential information residing in the molecules and structures of living systems is significant or meaningful to the system. Sets of random sequences or identically repeated sequences, for example, would be expected to contribute little or no useful information to a cell. This issue of quantitation of information is important since the ebb and flow of biologically significant information is essential to our quantitative understanding of biological function and evolution. Motivated specifically by these problems of biological information, we propose here a class of measures to quantify the contextual nature of the information in sets of objects, based on Kolmogorov's intrinsic complexity. Such measures discount both random and redundant information and are inherent in that they do not require a defined state space to quantify the information. The maximization of this new measure, which can be formulated in terms of the universal information distance, appears to have several useful and interesting properties, some of which we illustrate with examples.

 \* *Corresponding author,* dgalas@systemsbiology.org *or* galasd@battelle.org
 + *Current Address*: Department of Chemical and Biomolecular Engineering &
                      Institute for Genomic Biology, University of Illinois, Urbana-Champaign

## Introduction

A living system is distinguished from most of its non-living counterparts by its storage and transmission of information. It is this biological information that is the key to biological function. It is also at the heart of the conceptual basis of systems biology. Bio-information resides in digital sequences in molecules like DNA and RNA, 3-dimensional structures, chemical modifications, chemical activities, both of small molecules and enzymes, and in other components and properties of biological systems, but depends critically on how each unit interacts with, and is related to, other components of the system. Biological information is therefore inherently context-dependent which raises significant issues concerning its quantitative measure and representation. An important issue for the effective theoretical treatment of biological systems is: how can context-dependent information be usefully represented and measured? This is important both to the understanding of the storage and flow of information that occurs in the functioning of biological systems and in evolution.

There have been several attempts to address this problem for biological complexity. Gell-Mann has stated one part of the question clearly and suggested an approach to answering it [1, 2]. Standish has also suggested that context-dependence is a critical problem for the understanding of complexity in general and has discussed algorithmic complexity and the invocation of context in terms of universal Turing machines as an approach to its solution [3]. Adami and Cerf have formulated a solution to their particular formulation of the problem for macromolecular sequences by defining an imaginary ensemble of sequences and appending to that the context for interpretation as an explicit set of constraints as to which positions are conserved and which are "random" [4]. There have also been attempts to grapple with the issue of structural complexity, related to the problem defined here: most notably the idea of "thermodynamic depth" of Loyd and Pagels [5], and the idea of "causal states" of Crutchfield and Shalizi [6]. We take a different approach here, which is to deal directly with the complexity of sets of bit-strings. This provides a general approach to context dependence, which should be applicable to many problems that depend on the complexity of a system, and provide computational tools applicable to real biological problems. We find that the direct approaches provided by the powerful concepts of intrinsic complexity pioneered by Kolmogorov [7], Chaitin [8], and Solomonoff [9, 10], and extended by Li, Vitanyi, Gacs and others [11-14] can be applied to this problem in a particularly simple way. Furthermore, the construct of an information manifold derived from the demonstration of a well-defined metric, an "information distance" [11, 13], lends itself well to a class of set-based information measures, including mutual information, and illuminates its meaning.

Multiple types of information can be represented by bit-strings. Shannon information [15], devised to deal with communications channels, is defined in terms of the ensemble of all possible messages or bit strings (a state space), a fundamentally probabilistic definition which is related to physical entropy. Kolmogorov-Chaitin-Solomonoff (KCS) information (we use the terms "information" and "complexity" interchangeably here), on the other hand, is intrinsic to the object. It is based fundamentally on the difficulty of describing the object – the more difficult it is to describe by a computational process, the more information is present [16]. In the KCS conception the definition of the information in a bit string is the length of the shortest computer program (on a universal Turing machine) whose output is this bit string. This definition, often called "Algorithmic Information Content" (AIC), while elegant, and conceptually powerful, is not computable. It has become clear, however, that compression algorithms can be used to estimate the KCS information in a bit string in several ways [11, 17] - the maximum possible compression gives the best estimate for the KCS information of a bit string, which can be estimated by a suitable compression algorithm (like the lossless Lempel-Ziv or Kieffer-Yang algorithms.) This idea, which is a practical implementation of the abstract idea of "Computable Information Content" (CIC), enables the practical use of KCS complexity. This alternative to probabilistic, statistical approaches allows consideration of information absent any knowledge of the ensemble of all possibilities from which the object is drawn, knowledge which is often impractical and presents a number of difficulties in biology. It is problematic when considering sequences of macromolecules in that the state space is usually defined by a construct of questionable significance, like "the ensemble of all possible sequences" or "all possible functional mutant forms of a protein sequence" [4]. This powerful idea of the "intrinsic complexity" of a string, in contrast to a probability-based measure, is the hallmark of the Kolmogorov or KCS complexity. Whenever the sample space and probabilities are well defined, useful calculations can be done using Shannon information, but this is often not the case in biology. Another advantage of the KCS information is that it can be viewed as an absolute and objective quantification of the information in a specific string or object. Absolute information content of individual objects, rather than the average information to communicate objects produced by a random source (the key concept of Shannon information theory) is clearly preferable.

From what we know of biological function at the molecular level, the interactive, highly connected networks with systems-like behaviors suggest to us that any measures that don't take this kind of context into account will be less than useful in accounting for biological information content. Protein-protein interaction networks, metabolic networks and gene regulatory networks are examples of the remarkable complexity of biological networks, and indicative of the importance of context. It is useful to approach the context question, however, by considering two "paradoxes" that illuminate the problem of information in biology.   A "random" bit string, $r$, has maximum KCS complexity, and therefore contains the most information, for a string of this length.   Another way of looking at the information in a random string, though this is a difficult issue [12, 14], is that by definition it is "incompressible" and can only be represented by a string of approximately its own length, $L(r)$;  i.e.  for a minimal description $C(r)$, $L(C(r)) \approx L(r)$.   The proper definition of "randomness" actually makes use of this notion [18].   In spite of this way of measuring information content, a random sequence, however, is devoid of useful information. This is a problem that Kolmogorov grappled with and responded by defining his *structure function* (attribution in [19].)  A random sequence has essentially no biological information (e.g. a random protein sequence has essentially no functional use) - the cell containing this sequence is therefore not more complex than the one without it, and we should be counting its contribution to the complexity of the set of information in the cell as <u>zero</u>.   Gibberish doesn't help with any biological process, to paraphrase Gell-Mann [2].  This is the first "paradox".

Consider on the other hand, adding not a random sequence but one that exactly matches a sequence that already exists in the set.  Using a context-free measure, that does not consider the other content of the set, this should add an amount of information equal to the existing sequence that it matches. The duplication of existing information in the set (e.g. like the exact duplication of a gene) adds less  new information to the set than the original, duplicated bit string, however. This is the second "paradox", though it seems a weaker one.

A good measure of information should therefore discount the addition of either random or pre-existing information to resolve these paradox-like conflicts. Both are dependent on the <u>relationship</u> of the information to other information in the cell.  Since biological function depends on relationships within the system, a measure of complexity, and a good definition of biological information, must account for relationships and context.   Our approach to the problem of biological information, therefore, will be to construct a measure of information in a set of bit strings, since this is general enough to deal with most problems.

Consider then an unordered set of *N* bit-strings, $S = \{x_i, i = 1, \ldots N\}$. The information in each of the strings individually may be described as KCS information or complexity, *K(x$_i$)*. Our biological measure of information must also reflect, however, the relationships to other strings in the set because there is some shared information that determines "function". For example, there is some information in the structure and sequence of one protein in the structure and sequence of any protein that interacts with it. Taking all interactions and structures into account, therefore, a protein interaction network is remarkably rich in information. There is also information about a metabolic pathway in an enzyme that catalyzes a reaction in the pathway and is coupled to it by product inhibition, or affected by the binding of other proteins. There are many other examples of biological context that contain information. Our challenge is to quantify it in a useful way. The <u>relationship between strings</u>, then, is what we characterize to begin tackling the biological information problem.

## Context-dependent Information Measures

A useful measure of information in one string, since it must include contributions from the other strings in the set, must be a function of the entire set. With this in mind we can approach the problem by defining a measure with a number of properties that we can specify.

1. A <u>random string</u> adds <u>zero</u> information to the set.
2. An <u>exactly duplicated string</u> adds little or nothing to the overall information in the set.
3. The measure includes the information content of the strings individually as well as the <u>information contained in the relationships with other members of the set.</u>

The second criterion is imprecise, but important, and one that we will return to discuss in more detail. The simplest and most direct measure of the information in a set might be the simple sum of the information in the individual bit-strings.

$$\Theta \equiv \sum_{i=1}^{N} K(x_i) \tag{1}$$

Clearly, this measure fails to satisfy our criteria and does not have any of the desired properties. We can, however, modify the definition in a simple way. The contribution of each string, $x_i$, is modified by a function, $F_i$, that depends on the <u>entire set</u>, and is therefore inherently context-dependent:

$$\Psi(S) \equiv \sum_{i=1}^{N} K(x_i) F_i(S). \qquad\qquad (2)$$

If $F_i=1$ for all $i=1,....N$, then $\Psi$ reverts to $\Theta$ and there is no context dependence, but if we take into account the dependence of $F_i$ on the relationships between the $i$th string and the others in the set we can construct a function that satisfies the three criteria above. Before we make that construction we need to introduce another useful measure. The informational relationship between strings has been well studied and a particularly useful measure exits, that of "universal information distance". We consider this function over the pairs of members of the set in the next section.

**Universal Information Distance and Set Information**

A normalized information distance function between two strings, $x$ and $y$ has been defined by Li, Vitanyi et al. [11]. It has been shown to define a metric in that it satisfies the three criteria: identity, the triangle inequality, and symmetry. This normalized distance measure

$$d(x,y) \equiv \frac{\max(K(x \mid y), K(y \mid x))}{\max(K(x), K(y))} \qquad\qquad (3)$$

takes values in [0,1]. For a set of strings the metric defines a space with a maximum distance of 1 between strings in all dimensions – in general a set of $N$ strings determines an $(N-1)$-dimensional space. A non-normalized information distance, defining a metric had been previously proposed by Bennett et al. [13], but for reasons articulated in [11], including the diverse lengths of strings of potential interest, is inadequate for most strings of interest to us. The normalized information distance is a powerful measure of similarity in that, as Li, Vitanyi and colleagues have shown, it is truly universal as it discovers all computable similarities between strings [11]. We can use the important "additivity of complexity" property which was proved (in a difficult proof) by Gacs [14]: $K(x,y) = K(x) + K(y \mid x^*) = K(y) + K(x \mid y^*)$ (where the equal sign means "to within an additive constant" in this equation, and $K(x,y)$ is the joint Kolmogorov complexity of $x$ and $y$) to rewrite eqn. 3 as

$$d(x, y) = \frac{K(x, y) - \min(K(x), K(y))}{\max(K(x), K(y))} .$$

(4)

The symbol $y*$ indicates the shortest program that generates $y$, which then gives us the additive constant relation. If we simply use the string $y$ itself, the equality is true to within "logarithmic terms" [14, 20, 21]. The difference is crucial for some applications, but is not important in the use we make of the relation.

Since the complexity of a finite string is a finite we can order the strings in the set by increasing complexity, and index them such that if $i > j$ then $K(x_i) \geq K(x_j)$. Thus, since the joint complexity is symmetric, we have

$$d(x_i, x_j) = \begin{cases} \dfrac{K(x_i, x_j) - K(x_j)}{K(x_i)} = \dfrac{K(x_i \mid x_j)}{K(x_i)} & if \quad i > j \\[3mm] \dfrac{K(x_i, x_j) - K(x_i)}{K(x_j)} = \dfrac{K(x_j \mid x_i)}{K(x_j)} & if \quad i < j \end{cases}$$

(5)

The average distance between pairs of strings in the set, for example, can be calculated from the sum:

$$\sum_{i,j} d(x_i, x_j) = \sum_{i > j} \frac{K(x_i \mid x_j)}{K(x_i)} + \sum_{i < j} \frac{K(x_j \mid x_i)}{K(x_j)}$$

(the larger $K$ is always in the denominator, and we ignore $i=j$ since $d(x_i,x_i)=0$). Using the symmetry of $d$ evident in the above expression the average can now be written as

$$\langle d \rangle_S = \frac{2}{N(N-1)} \sum_{i > j} \frac{K(x_i \mid x_j)}{K(x_i)}$$

(6)

A similar expression, the "complexity-weighted" average distance, $\Lambda_S$, reduces to a simple expression since, in a set ordered by increasing complexity, we have a very simple expression for the conditional complexity, $K(x_i \mid x_j) = d(x_i, x_j) K(x_i), \quad \forall i > j$. Thus

$$\Lambda_S \equiv \frac{2}{N(N-1)} \sum_{pairs} K(x_i) d(x_i, x_j) = \frac{2}{N(N-1)} \sum_{pairs} K(x_i \mid x_j) = \left\langle K(x_i \mid x_j) \right\rangle \qquad (7)$$

The average conditional complexity over the set is thus equal to the "complexity-weighted" average distance. The conditional complexity of each pair of strings is the larger of the complexities of the two times the distance between the strings in this space .

## **Set Complexity**

We turn now to the three criteria that must be satisfied for a function of the set, $F_i(S)$, described in eqn. 2. The first criterion is satisfied if a random bit string, $x_i$, added to the set $S$ makes $F_i(S) = 0$. If in doing this we ensure that $F_i(S)$ depends on the entire set then criterion 3 is also satisfied. These two criteria can be fulfilled by expressing $F_i(S)$ in terms of the universal information distance, which leads to a simple definition of set complexity.  Clearly, if a bit string $x_i$ is random, then (for a sufficiently long string) the distance in the information manifold from <u>any</u> string is just 1, the maximal distance.  Thus, criterion number 1 can be met in general by any expression that includes a sum over the entire set:

$$F_i(S \bigcup x_i) = G_i(S \bigcup x_i) \sum_j f(d_{ij}) \qquad (8)$$

(the choice of a summation is somewhat arbitrary here) where $G$ is any positive finite function over sets of bit strings, $d_{ij}$ is shorthand for $d(x_i, x_j)$, and $f(x)$ is a function of the pair-wise universal distances, positive in (0,1), <u>with a zero at $x = 1$</u>.  The zero at the maximum distance is key.  This relationship applies to any set, $S$.  Criterion number 2, however, is a somewhat loose constraint. Let's consider it first by the case of adding an identical string to a set that consists of all identical bit strings only.  For such a set, S, then, the condition can be satisfied exactly if we use the summation in a similar way

$$F_i(S \bigcup x_i) = G_i(S \bigcup x_i) \sum_j g(d_{ij}) \qquad (9)$$

where $G$ is any finite function of the set, and $g(x)$ has <u>a zero at $x = 0$</u>.   What this constraint means for an arbitrary set (not only the set of all identical strings) is that the increase in complexity of the set saturates as the number of identical bit strings increases (it does not continue to increase with the size of the set.)    We can now define the set complexity, $\Psi$, in terms of the above functions $f$ and $g$ that satisfy all three criteria.  We simply set $G = 2/(N-1)$ for normalization, and multiply $f$ and $g$ to get

$$\Psi(S) = \sum_{i=1}^{N} K(x_i) \left\{ \sum_{pairs} \frac{2}{N(N-1)} f(d_{ij}) g(d_{ij}) \right\} \qquad (10a)$$

and since the simplest functions that satisfy these criteria  are   $f(d) = (1-d)$  and  $g(d) = d$ , we can appeal to parsimony and define the correspondingly simplest expression for set complexity:

$$\Psi(S) = \sum_{i=1}^{N} K(x_i) F_i(S), \quad \text{where}$$
$$F_i(S) = \frac{2}{N(N-1)} \sum_{pairs} d_{ij}(1-d_{ij}) \qquad (10b)$$

where the sum over pairs is consistent with the ordering of the set. Finally, we can generalize this function, call it $\tilde{\Pi}$, to any that have zeros at 0 and 1 so that it satisfies all the criteria we specified:

$$\Pi_i(S) = \frac{2}{N(N-1)} \sum_{\alpha=1}^{\infty} \sum_{\beta=1}^{\infty} a_{\alpha\beta} \sum_{pairs} d_{ij}{}^{\alpha} (1-d_{ij})^{\beta}$$
$$d_{ij}{}^{\alpha}(1-d_{ij}^{\beta}) = \sum_{n=0}^{\beta} \binom{\beta}{n}(-d_{ij})^{n+\alpha} \qquad (11)$$

This is a context-dependent measure of set complexity which we will consider in this paper only in its simplest form:  eqn 10b.  In addition to this form there are other simple ones such as $d_{ij} \ln d_{ij}$ , or  $(1-d_{ij}) \ln(1-d_{ij})$ .

## Relationship to Mutual Information

The idea of mutual information is a central concept in understanding the sharing of information between two objects, in our case bit strings. Mutual information quantifies the information in string $y$ about string $x$, and is symmetric.  These properties can be defined in both algorithmic (individual) and probabilistic terms [21], and the algorithmic concept represents a significant sharpening of the probabilistic notion.  The mutual information, $I$, between two strings, $x$ and $y$, can be defined in terms of complexity in the notation of Gacs et al. [21], using the additivity relation (see first section).

$$I(x : y) = K(x) - K(x \mid y^*)$$
$$= K(x) + K(y) - K(x, y) \qquad (12)$$

(the second equality is again within an additive constant, and the symmetry has this character as well) Since the distance between $x$ and $y$, if $K(x) > K(y)$, can be written as $\quad d(x, y) = \dfrac{K(x \mid y)}{K(x)}$ we can express the mutual information (to within the accuracy pointed out in [14, 21]) in terms of this distance, which gives us

$$I(x : y) = K(x) - K(x)d(x, y) = K(x)(1 - d(x, y)) \qquad (13)$$

where the same ordering constraints apply. A definition of set complexity using mutual information, of course, contains context information, and so it is useful to make such a definition by constructing such a function $F$.  In fact, this is what is represented in eqn. 8 with $f = (1-d)$. Thus,

$$\Phi \equiv \frac{2}{N(N-1)} \sum_{pairs} K(x_i)(1 - d_{i,j}) \qquad (14)$$

(Again the sum over pairs assumes that $K(x_i)$ is the larger of the complexities of the two strings of each pair.)  This *mutual information* sum, $\Phi$, actually resolves one of our "paradoxes". It discounts random sequences entirely, but the identical sequence problem remains unsolved.  The mutual information measure, when expressed in the metric space, is close to our constraint-based measure $\Psi$, (having a zero at $d=1$), but it is insufficient to resolve the second "paradox" since there is no zero at $d=0$.  Nonetheless, the relation between $\Phi$ and $\Psi$ is illustrative of a large class of set information measures defined entirely in terms of pair-wise distances in an information manifold.  This represents a large class of measures, as shown in eqn. 10a.    To illustrate explicitly the first few of these "statistics" (indicating the normalization factor by $\xi(N) = \dfrac{1}{N-1}$ ) .

$$M(S) \equiv \sum_i K(x_i) F_i(S) \qquad (15)$$

$$F_i(S) = 1 \qquad\qquad M = \Theta, \text{ simple sum ( } \underline{no} \text{ context dependence)}$$

$$F_i(S) = \xi(N) \sum_{j<i} d_{ij}, \qquad\qquad M = \Lambda, \text{ weighted average distance,}$$
$$\text{or average conditional complexity}$$
$$\text{(context-dependence)}$$

$$F_i(S) = \xi(N) \sum_{j<i} (1 - d_{ij}), \qquad\qquad M = \Phi, \text{ mutual information (context-dependence)}$$

$$F_i(S) = \xi(N) \sum_{j<i} d_{ij}(1 - d_{ij}) \qquad\qquad M = \Psi, \text{ satisfies constraints ( context-dependence)}$$

This set of measures, focused on the properties of the metric function, $F$, is clearly representative of a much larger set of interesting set functions illustrated in eqn. 11. All but the first of these functions give a context dependent measure. The last one is the simplest possible function with zeros at 0 and 1, and represents our chosen measure function. In general, these information measures differ by the weightings given to the distribution of relative string complexities in the set. A simple relationship between these functions is evident from eqn. 15: $\Phi(S) = \Theta(S) - \Lambda(S)$ . There is another that is less obvious that we now consider in the next section.

## "Mean field" Approximation and Fluctuations in the Information Manifold

Since we can express our information measure in terms of information distances we can usefully examine the relationship between the above measures and the variations in a set of strings in the same metric terms. We can relate $\Psi$ to $d$ using the relations in eqn 15.

$$\Psi(S) = \xi(N) \sum_{pairs} K(x_i)(d_{ij} - d_{ij}{}^2) = \Lambda - \xi(N) \sum_{pairs} K(x_i) d_{ij}{}^2 \quad (16)$$

Then with a term that represents the underline{complexity-weighted distance variance} of the set, we have.

$$\Psi(S) = \Lambda - \Lambda^2 + \Lambda^2 - \xi(N) \sum_{pairs} K(x_i) d_{ij}^{\ 2}$$

$$= \Lambda(1-\Lambda) - \Delta^2 \qquad\qquad (17)$$

$$\Delta^2 \equiv \xi(N) \sum_{pairs} K(x_i) d_{ij}^{\ 2} - \left( \xi(N) \sum_{pairs} K(x_i) d_{ij} \right)^2$$

The above expression for $\Psi$ looks intriguingly like a "mean field" term plus a fluctuation term. The "mean field term" is essentially a reflection of the form of the distance-function in Eq. 10b. The fluctuation term, $\Delta^2$, measures the degree of deviation of the complexity distribution of the set, and any deviation reduces the overall set complexity. We can make this more precise. The mean-field-like approximation ($\Psi_{mf}$, the mean-field approximation to $\Psi$) simply sets $\Delta$ to zero. Then we have the complexity-weighted distance average or the average conditional complexity, $\Lambda$ as the key variable: $\Psi_{mf} \propto \Lambda(1-\Lambda)$. We can describe the mean-field in terms of this statistic. Further, if the bit strings were uniformly spaced at distance $d$, then $\Delta^2$ and $\Psi$ would simplify (carefully accounting for the numbers in the pairs sum):

$$\Delta^2_{uniform} = d^2 \sum_i K(x_i) - \left( d \sum_i K(x_i) \right)^2$$

$$\Psi_{uniform} = \Lambda(1-\Lambda) - d^2 \Theta(1-\Theta) \qquad\qquad (18)$$

but since $\Lambda_{uniform} = d\Theta$, a very simple expression (also obvious from Eq. 10b) emerges:

$$\Psi_{uniform} = d(1-d)\Theta \qquad\qquad (19)$$

If we make the further simplifying assumption that the uniform distance and the total of the complexities can be varied separately, then it is clear that the maximum of $\Psi$ occurs when $d = \frac{1}{2}$.

## Computational estimation and application of set-based information

The set complexity based on inherent KCS complexity of strings has many advantages, as discussed, but the definition of set complexity (eqn. 12) is inherently incomputable. Thus we need to introduce a computational approximation for this quantity. Data compression has been used to make this kind of approximation (see [22] for a comparison of several approaches). It has been shown that the universal information distance can be approximated using the Normalized

Compression Distance (NCD) and this is comparable to a different approximation, <u>Universal Compression Distance</u> (UCD). We will use NCD here

$$d_{NCD}(x,y) = \frac{C(xy) - \min\big(C(x), C(y)\big)}{\max\big(C(x), C(y)\big)} \qquad (20)$$

where $C(x)$ is the compressed size of string $x$ and $xy$ is a concatenation of strings $x$ and $y$ [17]. This approximation is based on the estimation of Kolmogorov complexity using a real compression algorithm, and makes use of the additivity property (see eqn. 4). By replacing the KCS complexity $K(x)$ by a computational approximation $C(x)$, the set complexity can be defined simply as

$$\hat{\Psi}(S) = \xi(N) \sum_{pairs} C(x_i) d_{NCD}(x_i, x_j)\big(1 - d_{NCD}(x_i, x_j)\big) \qquad (21)$$

The compressed size $C(x)$ of a string $x$ is an upper-bound for the Kolmogorov complexity $K(x)$.

Even though the NCD can be applied to approximate the universal information distance with remarkable success, one issue is that the range of NCD may be smaller than [0,1]. In some cases, the estimate of NCD can even take on values larger than one [17]. As our measure of set complexity is based on the assumption that the distance between two identical strings as well as between two random strings approaches zero, problems in the estimation of set complexity can arise, since the errors in the NCD accumulate in the sum of distances in Equation (21). We can address this problem by introducing scaling factors for the computed NCD values, and normalizing the obtained distances to the [0,1]-interval. These scaling factors can be obtained by computing the minimum and maximum observable distance for a given set of data. The minimum distance is obtained by computing the distance between two identical strings; that is, for each string in a set, compute the distance to itself and select the one that has the smallest value. The maximum distance is obtained by comparing random strings; that is, for a set of strings, permute the strings and find the maximum distance among all permuted strings within the set.

In order to study the behavior of the estimated set complexity, $\hat{\Psi}$, we considered a set of 25 random, but identical, binary strings of length $L = 1000$ and used the familiar *gzip* compression algorithm to estimate the Kolmogorov complexity (this is based on the Lempel-Ziv algorithm).

We introduce noise by randomly perturbing one bit at a time in each string. The set complexity for different amounts of noise is shown in Figure 1. It can be seen that as the noise is introduced, the set complexity increases until the amount of noise exceeds a certain value. As the individual strings start to loose common structure, the set complexity begins to decrease as the set becomes more and more random. Due to the approximation issues discussed above, the set information does not go to zero for either the identical or randomized sets. There are two sources of error in the compression approximation: 1. the estimates of randomness are inherently poorer the shorter the length of the strings – specifically the distance between finite bit strings never goes to 1 and the accumulated error can be substantial; and 2. the computational issues mentioned above (see [17, 22]).

We can actually study the errors in our approximation by computing set complexity under conditions whose outcome we know *a priori*. An experiment of this kind can be defined as follows. Start from a set of all identical strings. As discussed earlier, the set complexity for this set should be zero. Then, replace one string at a time by a completely random string. As a random string does not contribute to the complexity since it should be distant from all others ($d \approx 1$), the set information should remain zero. This can be repeated for all strings in a set, leading to a set of random strings whose set information should also equal zero. Thus, with this process we can move from a set of identical strings to a set of random strings, generating a series of sets that should all have set information of zero if our approximation were exact. The result of the computational approximation of this process is shown in Figure 2. It can be seen that, in practice, the estimated complexity of a set of random strings is larger than the complexity of a set of identical strings. This is not unexpected, as for finite, randomized strings, and with the *gzip* approximation, the residual mutual information estimate is clearly not zero. Overall the estimate of Ψ has a difference of about 1.8-fold at the extreme ends of these test sets. This enables us to estimate the computational error in the estimated informational distances. Each "random" string has a calculated distance from the others of about 0.92 on average. We can then refine our calculations take this average error into account, and use the randomized strings to adjust the estimates of distance in our computational estimation. The deviation in Figure 2 was so used to adjust the process shown in Figure 1, resulting in the normalized set complexity estimate shown in Figure 3. With this adjustment there is no significant difference in complexity between a set of identical strings and a set of random strings.

## Criticality in the Dynamics of Boolean Networks

As another application of our set complexity measure, we decided to examine the amount of information contained in the state dynamics of a model class of complex systems that can exhibit ordered, chaotic, and critical dynamics. For simplicity we consider random Boolean networks (RBNs), which have been extensively studied as highly simplified models of gene regulatory networks [23, 24] and other complex systems phenomena [25].

In a Boolean network, each node is a binary-valued variable the value of which (0 or 1) is determined by a Boolean function that takes inputs from some subset of nodes, possibly including the node itself. In the simplest formulation, all nodes are updated synchronously, thereby generating trajectories of states, where a state of the system at a particular time is an *n*-length binary vector containing the values of each of the *n* nodes in the network. Boolean network models for several biological gene regulatory circuits have been constructed and shown to reproduce experimentally observed results [26-29].

In an RBN, each of the *n* nodes receives input from *k* nodes (determined by the random structure of the network) that determine its value at the next time step *via* a randomly chosen Boolean function assigned to that node. The output of each such function is chosen to be 1 with probability *p*, which is known as the *bias* [30]. Thus, the parameters *k* and *p* can be used to define ensembles of RBNs. In the limit of large *n*, RBNs exhibit a phase transition between a dynamically ordered and a chaotic regime. In the ordered regime, a perturbation to one node propagates on average to less than one other node during one time step, so that small transient perturbations to the nodes die out over time. In the chaotic regime, such perturbations increase exponentially over time, since a perturbation propagates on average to more than one node during one time step [25]. Networks that operate at the boundary between the ordered and the chaotic regimes have been of particular interest as models of gene regulatory networks, as they exhibit complex dynamics combined with stability under perturbations [24, 31-33].

For network ensembles parameterized by *k* and *p*, an order parameter called the average sensitivity [34], given by $s = 2kp(1-p)$, determines the critical phase transition in RBNs by specifying the average number of nodes that are affected by a perturbation to a random node. Thus, the ordered regime corresponds to $s < 1$, the chaotic regime to $s > 1$, and the boundary at $s = 1$ defines the point of the phase transition. The average sensitivity corresponds to the well-known probabilistic phase transition curve derived by Derrida [35]. It is also easily computable for a particular network given its set of Boolean update functions. The logarithm of the average sensitivity can be interpreted as the Lyapunov exponent [36]. Thus, by tuning $k$ and $p$, networks can be made to undergo a phase transition.

Networks that operate close to the critical regime can exhibit the most complex dynamics, as compared to ordered or chaotic networks. Indeed, ordered networks give rise to simple state trajectories, meaning that the states in a trajectory are very similar, periodic, or quasi-periodic and often identical due to the "freezing" of a large proportion of nodes in the network. Chaotic networks, on the other hand, tend to generate "noisy" state trajectories that in time become indistinguishable from random collections of states when the parameters are deep in the chaotic regime. In both these regimes, the set complexity of a randomly chosen state trajectory might be expected to be small, since it should contain a set of nearly identical or nearly random states.

We examined this question by applying our NCD-based estimate of set complexity to state trajectories generated by ensembles of random Boolean networks operating in the ordered, chaotic, and critical regimes. Specifically, we have set the connectivity to be $k = 3$ and tuned the bias $p$ in increments of 0.01 so that the average sensitivity, $s$, varies from $s < 1$ (ordered) to $s > 1$ (chaotic). For each value of $s$, 50 random networks (number of nodes, $n = 1000$) were each used to generate a trajectory of 20 states, after an initial "burn in" of running the network 100 time steps from a random initial state in order to allow the dynamics to stabilize (i.e., reach the attractors). We collected these 20 network states into a set $S$ for each network of the ensemble and calculated $\hat{\Psi}(S)$ for each. Figure 4 shows the average $\hat{\Psi}(S)$ over the 50 networks as a function of the average network sensitivity $s$ (plotted a function of $\lambda = log\ s,$ the Lyapunov exponent).

It is clear that networks that are operating close to the critical regime have the highest average set complexity of their state trajectories. In addition, the variability of the set complexity is also highest near the critical regime, indicating that critical (or near critical) networks are capable of exhibiting the most diverse dynamics. When networks are deep in the ordered regime (far to the left), the average set complexity of their state trajectories is low and the variability is small. This can be explained by the relatively simple network dynamics, consisting mostly of frozen node states and nodes that exhibit short periodic dynamics. As networks become more chaotic, the states in the trajectories become more stochastic, resulting in a decrease in their set complexity. Our results clearly support the view that complex systems operating at or near criticality, a property that is believed to hold for living systems, appear to exhibit the most informationally complex dynamics. The measure $\Psi$ seems very well suited to describing capturing this phenomenon near criticality.

### Context-dependent Information of Networks

In the spirit of grappling with context-dependence in biological applications, we apply our complexity measure to networks. First, we consider only undirected graphs with unweighted edges, represented as an adjacency matrix such that $A_{ij} = 1$ if an edge connects nodes $i$ and $j$, and $A_{ij} = 0$ if not. While there are numerous methods for representing the similarity between individual nodes, our objective here is to quantify the global complexity of the graph in a way that balances regularity with randomness as discussed above.

To use our measure, we must first define the information content, or complexity, of each node. Since we have not yet defined any other attributes to the nodes or edges, this must derive from the connectivity of each node. This is represented in the complexity of the bit string $x_i$, the $i^{th}$ row vector of the adjacency matrix. The set of complexities is { $K_i$ }. In the same way we take inter-nodal information distances, $\{d_{ij}\}$, to be dual to the mutual information between nodes, with $d_{ij} = 1 - w_{ij}$ (see eqn. 15.) The complexity of the strings can be calculated using the KCS approach. This is a case where we can take the relationships we have in the algorithmic formalism and define the measures, such as mutual information, in the probabilistic sense. The subtle relationship between the two approaches is extensively discussed in references [21] and , for example. It is important to note that the algorithmic formalism is more fundamental, but in this case, where we have a well defined state space, the quantities can be calculated using the familiar Shannon entropy and mutual information based on row vectors in the adjacency matrix. For each node we have:

$$K_i = -\sum_{a=0}^{1} p_i(a) \log_2 \big( p_i(a) \big), \qquad (22)$$

where we consider only whether two nodes are connected or not ($a$ takes on the values 0 or 1 – that is; the alphabet describing the connections is binary) : $p_i(1)$ is the probability of the $i$-th node being connected and $p_i(0) = 1 - p_i(1)$ is the probability of it being unconnected (self connecting loops are not allowed.) The inter-nodal information distances are defined in a similar way. The Shannon mutual information between two variables, $a$ and $b$, is given by

$$I(a:a) \equiv H(a) - <H(a\,|\,b)>$$
$$= H(p(a)) - \sum_b p(a,b) H(p(a\,|\,b)) \tag{23}$$

where $H$ is the Shannon information and $<H(a|b)>$ is the average conditional information. (The second line of eq. 23 indicates the dependence of these quantities on the probability distribution of the variables $a$ and $b$.) Thus,

$$d_{ij} = 1 - \sum_{a,b=0}^{1} p_{ij}(a,b) \log_2 \left( \frac{p_{ij}(a,b)}{p_i(a)\,p_j(b)} \right) \tag{24}$$

Here $p_{ij}(a,b)$ is the joint probability of nodes $i$ and $j$ being related to a third node with value *(a,b)*, so the probabilities measure the relative prevalence of pattern of connectivity; e.g. $p_{ij}(1,1)$ is the probability of both being connected to another node. By taking logarithms of base 2 here, both $K_i$ and $d_{ij}$ can be normalized to the interval [0,1]. We can now apply eqn. 11 directly to compute the complexity $\Psi$, of the network. It is simple to show that for this measure (with binary value connections) a graph and its conjugate have equal complexity.

As we found above, the maximal value of $\Psi$ arises when all $d_{ij} = d = 1/2$ and is proportional to a uniform nodal information content $K$. With our normalized formulation this corresponds to all $K_i = 1$. From the "mean-field" approximation we can intuit that the maximally complex graph will have minimal variation in both single-node information and mutual information between nodes. This suggests that a somewhat uniform degree distribution corresponds to maximal complexity. We also can expect the degree distribution to be centered at $N/2$, since according to equation (23) nodes with equal numbers of neighbors and non-neighbors have maximal information content ($K = 1$ when $p_1 = 0.5$). However, it is important to note that perfect uniformity in degree distribution will actually lead to low complexity due to topological redundancy. For example, the union of two complete graphs ($K_{N/2} \cup K_{N/2}$) (or almost equivalently its conjugate, the complete bipartite graph $K_{N/2,N/2}$) generates a very low $\Psi$ (Figure 5A). A few edge rearrangements, however, that disrupt the uniformity transform such graphs into highly complex networks, as shown in Figure 5B.

Thus, for undirected, unweighted, Erdos-Renyi random graphs we find that maximal complexity arises from nearly bimodular or near-bipartite graphs. These graphs appear to balance the requirement of maximal complexity for each single node with the requirement of uniform mutual information between all node pairs.   This suggests that modular graph architecture adds information content, although it remains to determine how this finding translates to other classes of graphs (directed, weighted, scale-free, multiple edge types, etc.).   Since biological networks appear to be rather modular, this is an interesting correlation.    Since the countervailing requirements of maximum complexity for each node and high, uniform mutual information are balanced when we attempt to maximize $\Psi$, and since these two requirements are reminiscent of observed properties of biological networks, we expect that $\Psi$, or a closely related function, has strong biological meaning for networks. If we imposed other constraints on the network (e.g. functional constraints, specific motifs, a specific growth and evolution process) and maximize $\Psi$ with those constraints, $\Psi$ may take on a clear meaning.   While this is an important issue to explore it is beyond the scope of the present paper.

If we consider more realistic networks where different characters of the edges are important (different edge types, or "alphabets"), we will increase the descriptive alphabet of the edges beyond binary, identifying distinct types of node interactions.  For equations 22 and 24 the sums will now extend over the full alphabet describing those edge types.  We have used such an extended alphabet in our analysis of genetic interactions for which there are many possible types of interactions that can be usefully distinguished. In that case the problem of classification of interaction types corresponds to an optimization of $\Psi$ by alphabet reduction (Carter, Galitski and Galas, in preparation).

It is worth noting that the problem of maximizing $\Psi$ by reducing the alphabet size for networks, as just described, is an example of a very general problem – that of balancing the simplicity of the descriptors of relationships with the complexity being described (not unlike the problems for which the "information bottleneck method" was devised [37].)

### Discussion

In this paper we have used the intrinsic information concepts of Kolmorgorov-Chaitin-Solomonoff complexity to construct a simple measure for set-based information that provides a theoretical foundation for dealing with context-dependent biological information. Over the past 50 years or so the theoretical foundations have been well laid for defining the absolute information content of an individual object, and the underpinnings of the ideas of randomness and of probability theory that began much earlier. This is clearly the best way, in our view, to approach difficult information problems, like those we encounter in biological systems. We can avoid the difficulties of defining ensembles and probability distributions over sample spaces that are problematic to define, as is required for a Shannon-based approach, and we can bring to bear the gains in rigor and conceptual approach to information and complexity of the KCS insights. If the sample space and probability is naturally well defined, as in our network example, on the other hand, the context-dependent measure is amenable to Shannon-type calculation (see [20] for an excellent review of the relationships).

While it is often stated that biology is an information science [38-40], we are still far from having the tools to provide a general theoretical basis for dealing with it as such. It is difficult to overestimate the importance of dealing rigorously with context-dependence in biology. While it is often acknowledged as important it is often difficult and ungainly to deal with. Since the processes of natural selection are well known to be powerful sculptors of context dependence in biological systems, selecting complementary alleles of genes in a genome, for example, with ruthless efficiency, we expect a natural selection to be a prodigious generator of context dependence. We do find, in fact, that the context-based measure is particularly useful in deciphering gene interaction data (Carter, Galitski and Galas, in preparation.) Gene interactions, while of fundamental importance in biology, are only one example where context is expected to be highly significant.

It is important to note that context dependence, driven by natural selection, leads to a dynamic phenomenon long studied in biology, of which the allele interaction effect is but one example. The phenomenon of "symmetry breaking", which characterizes the loss of some symmetry or simplicity, the acquisition of new distinctions, is not fully understood or appreciated in complex systems. It is, however, widespread in biology – for example see [41, 42]. In order to deal with symmetry breaking generally and effectively we need to have a global formalism, and since it is the information that dominates our view, a global formalism for context-dependent information is just what's needed. We propose that the theory presented here is a beginning of the development of a class of tools for analyzing this aspect of biology. An important problem for future study then is that of describing an interactive dynamics for a biological process in the information manifold in terms of the complexities and distances. This holds the prospect for a deep understanding of the origins and evolution of biological broken symmetry in terms of biological information.

We began our construction by setting context-based constraints on an information measure for a set of bit strings, and we formulated two "paradoxes" for biological information which then guided us to find a measure that resolves them.  We found that a construct, based on the complexity of the bit strings in a set, can be expressed in terms of distances in an information metric space using the elegant and useful universal information distance of Li and Vitanyi [11], a normalized information distance.  The connection to information distances and the metric space of information (also called "universal similarity") provides a set of new tools for formulating problems in systems biology and evolution, as it promises to allow us to deal quantitatively with the ebb and flow of information in biology. Since this information is deeply context dependent there has been no consistent and rigorous way to grapple with these problems. Since biological information is inherently context-dependent, there have been significant issues with its quantitative representation using the usual information measures since both probabilistic and intrinsic approaches (Shannon and Kolmorgorov-Chaitin-Solomonoff) are inherently context-free in and of themselves.  Our formulation provides a solution to this problem.

Our approach leads to a general formulation allowing us to describe a very general class of information measures.  One of these proposed measures, $\Psi$, appears to solve some key conceptual difficulties of biological information – the "paradoxes" of the uselessness of both random and redundant information.  It is also the simplest form in the class of measures that will likely be useful in a variety of specific applications.  Our approach is quite distinct from previous work, like that of Adami and Cerf [4] that requires an ensemble of biologically functional examples and an explicit constraint representing the "environment".  While their approach can work for specific sets of functional molecules, like tRNA sequences, it is not useful for more complex problems, particularly when the ensemble is impossible to specify.   Nonetheless, it is clear that our theoretical construction is only a beginning. There are a number of remaining problems. One of these problems is the actual calculation of the KCS complexities.  There have been a number of important advances toward the estimation of complexity by the use of compression algorithms, but these methods are not always practical because of computational intensity, and they are inherently approximations whose accuracy is sometimes difficult to estimate.  It is clear that while the Shannon formulation is not particularly useful in many of the cases in which we are interested, it often does provide a practical approach to computation. Since it is clear that there is a direct correspondence between the probability-based and inherent complexity-based approaches (carefully reviewed by Grünwald and Vitanyi [20]) this provides a reasonable approach to practical computation in some cases, as in our description of network (graph) complexity. We are currently exploring the use of these techniques in a wide range of applications.

Among the major remaining problems we identify is the "encoding problem." While the representation of information in bit strings is a powerful and general approach, there remains the conceptual difficulty of encoding actual biological information in this representation (this is similar to the problem for macromolecular sequences that is "solved" in reference [4] by postulating a functional ensemble of examples. To describe, even under the simplest of assumptions, the information in the living cell that gives context to other pieces of information in the cell (all information that interacts with them) is a formidable challenge – this is what we call the "encoding problem."

The application of our approach to problems of complex systems analysis outside of the realm of biology should be a natural extension of the problems discussed here. One general problem of significant interest relates to the extension of the methods of "maximum entropy" and the ideas of "Occam's razor" [43] using context-dependent measures on sets. This extension should be straightforward, but the useful setting of and interpretation of the constraints is an interesting challenge. The potential similarity of these ideas to notions about perceived similarities between objects that are close in distance in a "psychology space" of some kind is also not lost on us [44]. In some real sense it is the distinction between objects based on the overall context of the set that determines the potential biological usefulness of the object – the analogy with the ideas from psychology is an interesting one, and not a little biological. Our purpose here is to lay the foundations of the quantitative theory, but we do not underestimate either the importance or the difficulty of this encoding problem, whose solution will be necessary for applications of our methods to real biological problems. We are currently working to extend our treatment to include a generalization and to grapple with this encoding problem explicitly by analyzing several model systems.

Acknowledgements: This work was supported by NSF FIBR grant EF-0527023, NIH P50 GM076547, NIH GM072855, the Battelle Memorial Institute, and the Academy of Finland, (supporting MN: project No. 120411 and No. 122973). NDP was supported by the Sam E. and Kathleen Henry postdoctoral fellowship from the American Cancer Society.

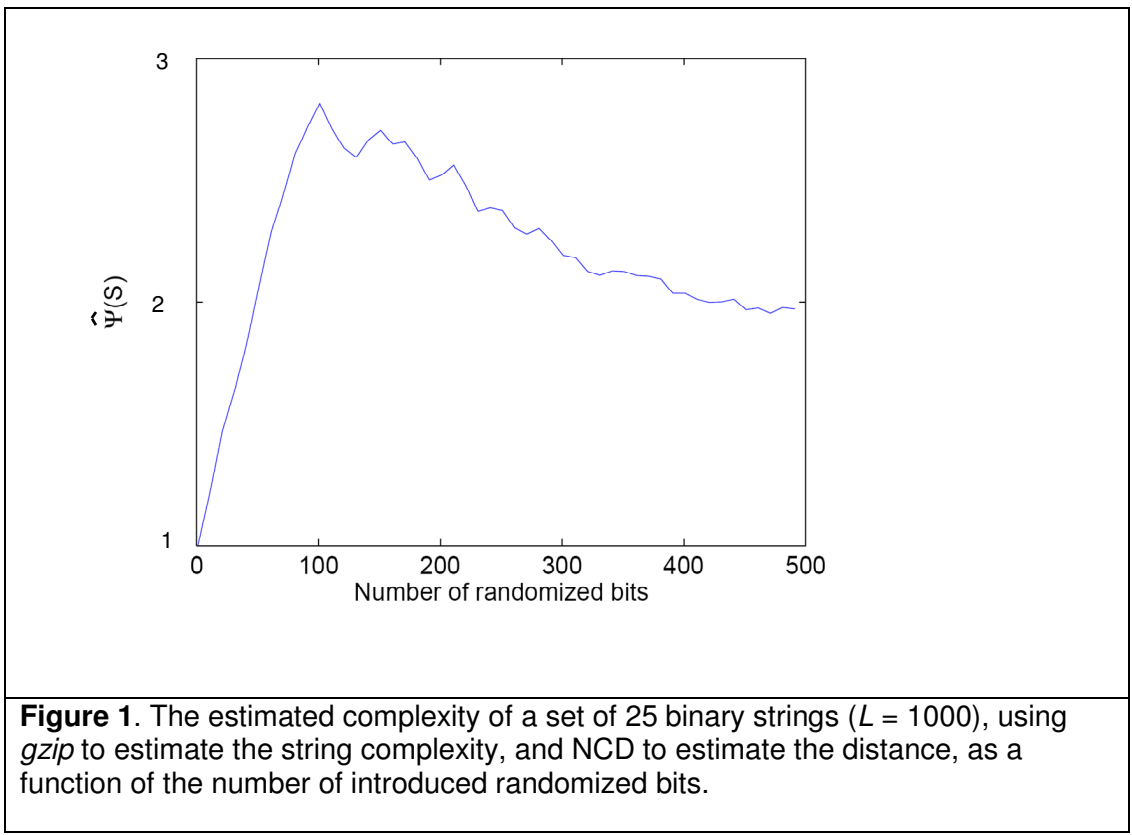

**Figure 1**. The estimated complexity of a set of 25 binary strings (*L* = 1000), using *gzip* to estimate the string complexity, and NCD to estimate the distance, as a function of the number of introduced randomized bits.

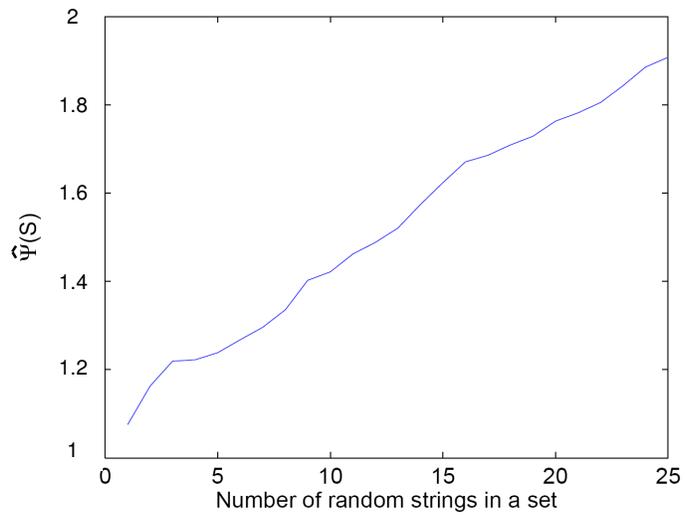

**Figure 2.** The estimated set complexity using the method employed in figure 1 (see text) as a function of the number of random strings substituted for identical strings in the set.

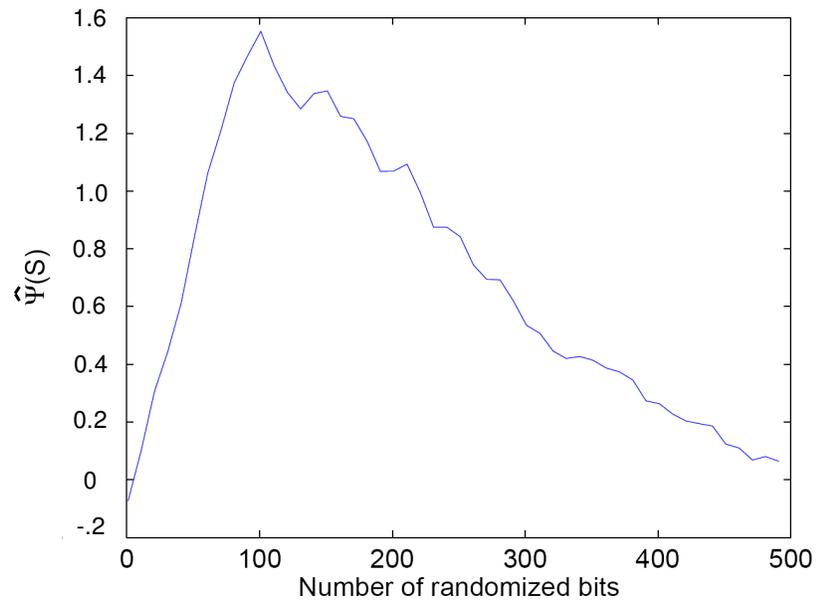

**Figure 3.** The set complexity of Figure 1 adjusted by the estimations of set complexity in Figure 2. As can be seen, the resulting set complexity of a set of identical strings is close to that of a set of random strings, as expected from such a measure.

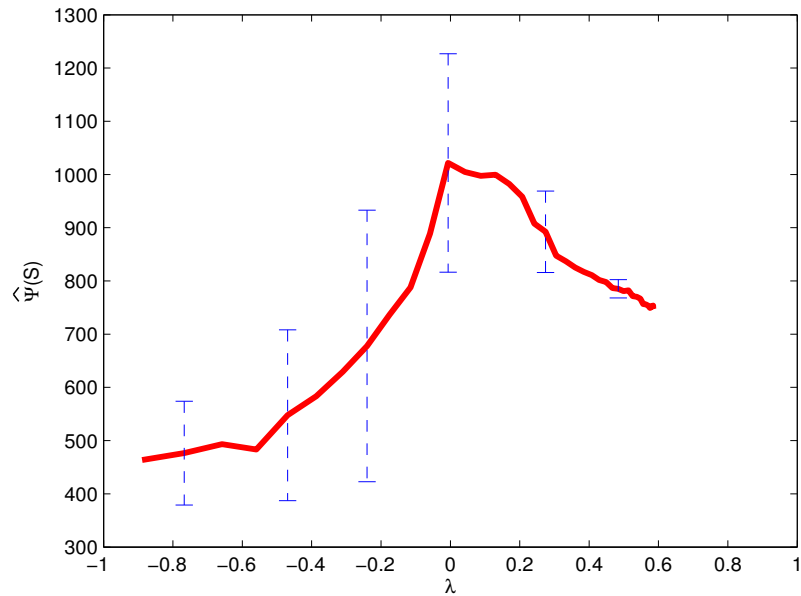

**Figure 4:** The average, estimated set complexity of random state trajectories as a function of the log of the average sensitivity λ, the Lyupanov exponent, generated by networks operating in the ordered (λ < 0), critical (λ = 0), and chaotic regimes (λ > 0). The bars show the variability (one standard deviation) of the estimated set complexity for 50 networks.

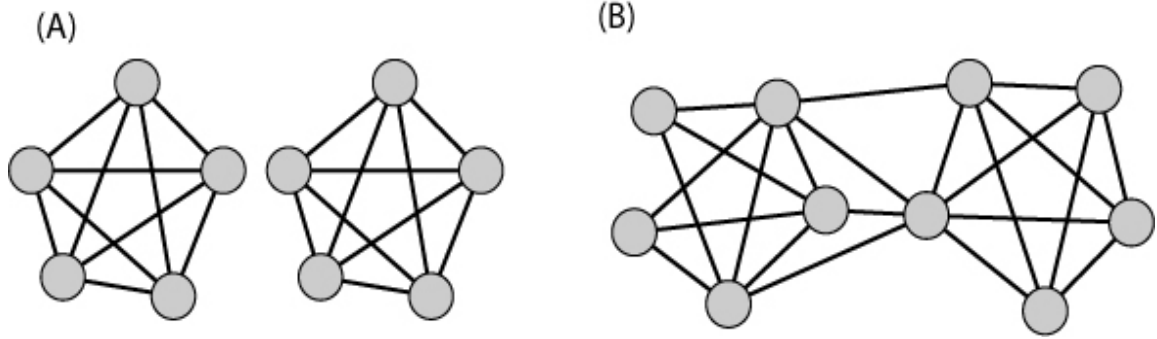

**Figure 5. Information content of two graphs with *N* = 10.** Graph (A) has a low information content: $\Psi_A = 0.2$. Graph (B), the maximally informative undirected, unweighted graph with $N = 10$, on the other hand, has a much higher information content: $\Psi_B = 1.9$.